\documentclass[acmlarge,screen,nonacm]{acmart}
\usepackage{booktabs} % For formal tables

\usepackage[ruled]{algorithm2e} % For algorithms

\SetAlFnt{\small}
\SetAlCapFnt{\small}
\SetAlCapNameFnt{\small}
\SetAlCapHSkip{0pt}

\setcopyright{none}
\renewcommand\footnotetextcopyrightpermission[1]{} % removes footnote with conference information in first column
\usepackage{empheq} 
\SetKwComment{Comment}{$//$\ }{}
\usepackage{tikz,pgfplots} 
\pgfplotsset{width=\linewidth, compat=1.9}
\usepackage{subcaption}
\usepackage{multirow}
\usepackage{dblfloatfix}

\usepackage{ifthen}

\newcommand{\makemath}[1]{{\ensuremath{#1}}}

\newcommand{\dt}{\makemath{h}} %time interval length
\newcommand{\timestep}{\makemath{t}} %time step

\newcommand{\pos}{\makemath{\mathbf{ q}}} % positions
\newcommand{\vel}{\makemath{\dot \pos}} % velocity
\newcommand{\acc}{\makemath{\ddot \pos}} % acceleration

\newcommand{\forcevec}{\makemath{\boldsymbol f}} % vector force
\newcommand{\forceextvec}{\makemath{\boldsymbol p}} % vector force
\newcommand{\forceconstraintvec}{\makemath{\boldsymbol c}} % constraint force

\newcommand{\solutionvec}{\makemath{\boldsymbol{\mathbf{b}}}} % solution vector to the conjugate gradient
\newcommand{\poscst}{\makemath{\mathbf{p}}} % positions on constraint space

\newcommand{\constraintNormal}{\makemath{\vv{\mathbf{n}}}} % constraint normal
\newcommand{\constraintFriction}{\makemath{\vv{\mathbf{f}}}} % frictional direction
\newcommand{\constraintDirection}{\makemath{\vv{\mathbf{c}}}} % uniformed direction

\newcommand{\mass}{\makemath{\mathbf{M}}} % Mass matrix
\newcommand{\damping}{\makemath{\mathbf{B}}} % Damping matrix
\newcommand{\stiffness}{\makemath{\mathbf{K}}} % Stiffness matrix
\newcommand{\contactJacobian}{\makemath{\mathbf{J}}} % Jacobian matrix
\newcommand{\contactJacobianMapping}{\makemath{\mathbf{H}}} % Isolated DOFs in Jacobian matrix
\newcommand{\contactJacobianDirection}{\makemath{\mathbf{C}}} % Transformation matrix for isolated dofs matrix and jacobian matrix

\newcommand{\systemMat}{\makemath{\mathbf{A}}} % Assembled system matrix
\newcommand{\delasus}{\makemath{\mathbf{W}}} % Deslasus operator
\newcommand{\delasusMapping}{\makemath{\mathbf{W_H}}} % Deslasus operator

\newcommand{\internalforcefunc}{\makemath{\mathcal{F}}} % Internal forces
\newcommand{\tr}{\makemath{\mathrm{^T}}}

\newcommand{\geomappingfunc}{\makemath{\mathcal{G}}} % Geometric mapping transform

\newcommand{\lamem}{\makemath{\mu}} % coefficient de LamÃ© 1
\newcommand{\lam}{\makemath{\mathbf{\boldsymbol{\lambda}}}} % constaint forces
\newcommand{\viol}{\makemath{\mathbf{\boldsymbol{\delta}}}} % violation of constaints
\newcommand{\fvec}{\makemath{\mathbf{f}}} % result b
\newcommand{\realspace}{\makemath{\mathbb{R}}}  %real space

\newcommand{\mappf}{\makemath{\mathcal{H}}} % mapping function of position
\newcommand{\comments}[3]{\ifthenelse{\equal{1}{#1}}{{#2}}{#3}}
\begin{document}
% Title portion
\title{An efficient implicit constraint resolution scheme for interactive FE simulations}

% DO NOT ENTER AUTHOR INFORMATION FOR ANONYMOUS TECHNICAL PAPER SUBMISSIONS TO SIGGRAPH 2019!
 \author{Ziqiu Zeng}
 \affiliation{%
   \institution{University of Strasbourg}
   \streetaddress{4 Rue Blaise Pascal}
   \city{Strasbourg}
 %  \state{VA}
   \postcode{67081}
   \country{France}}
 \email{zengziqiu1995@gmail.com}

 \author{Hadrien Courtecuisse}
 \orcid{0000-0001-6342-2284}
 \affiliation{%
   \institution{University of Strasbourg}
   \streetaddress{4 Rue Blaise Pascal}
   \city{Strasbourg}
 %  \state{VA}
   \postcode{67081}
   \country{France}}
 \email{hcourtecuisse@unistra.fr}

\begin{abstract}
\section{Abstract}
\label{sec: Abstract}
This paper presents a novel implicit scheme for the constraint resolution in real-time finite element simulations in the presence of contact and friction. Instead of using the standard motion correction scheme, we propose an iterative method where the constraint forces are corrected in Newton iterations. In this scheme, we are able to update the constraint directions recursively, providing more accurate contact and friction response. However, updating the constraint matrices leads to massive computation costs. To address the issue, we propose separating the constraint direction and geometrical mapping in the contact Jacobian matrix and reformulating the schur-complement of the system matrix. When combined with GPU-based parallelization, the reformulation provides a very efficient updating process for the constraint matrices in the recursive corrective motion scheme. Our method enables the possibility to handle the inconsistency of constraint directions at the beginning and the end of time steps. At the same time, the resolution process is kept as efficient as possible. We evaluate the performance of our fast-updating scheme in a contact simulation and compare it with the standard updating scheme.
\end{abstract}

\begin{CCSXML}
<ccs2012>
   <concept>
       <concept_id>10010147.10010371.10010352.10010379</concept_id>
       <concept_desc>Computing methodologies~Physical simulation</concept_desc>
       <concept_significance>500</concept_significance>
       </concept>
   <concept>
       <concept_id>10010147.10010169.10010170</concept_id>
       <concept_desc>Computing methodologies~Parallel algorithms</concept_desc>
       <concept_significance>500</concept_significance>
       </concept>
 </ccs2012>
\end{CCSXML}




\maketitle

\section{Introduction}
\label{sec: intro}

Simulating multi-object systems always receives strong interest in the computer graphics community.
Physics-based simulation is required to handle various materials containing rigid and soft solids.
For soft materials, dealing with the elastic functions in solid deformation leads to challenging problems.
The Finite Element Method (FEM) has been a gold-standard approach in this context \cite{Sifakis2012}.
On the other hand, coupling the simulation of different objects remains one of the other challenges in multi-object systems.
The interactions in physics-based animations have been intensively studied in many works. 
Readers can find a general introduction of contact simulation in \cite{Andrews2021}.

Simulating in real-time is essential for many tasks such as robot controlling \cite{Duriez2013a}, \cite{Baksic2021} and virtual surgeries \cite{courtecuisse2014real}.
In these tasks, the simulators are required to provide relevant information in real-time for users (or robots) to have efficient control and timely visual/haptic feedback to manipulators.
While computing the contact and friction responses, the time-stepping scheme is suitable for real-time simulations.
On this basis, constraint-based techniques have been prevalent for solving the complementarity problems as they guarantee non-interpenetration within a time step.
Another essential issue in FE simulations is that accuracy often requires sufficiently detailed discretization of solids, leading to costly computations for contact responses.
To address the issue, the methods using parallel architectures such as graphics processing units (GPUs) are developed, providing high computation speed while satisfying the requirement of accuracy.

% drawback of time stepping
% drawback of current method (H is only linearized once)
In practice, the constraints are discretized and linearized in time steps.
In many recent works, to simplify the solving process, such linearization is usually carried out once in each time step, making the constraint resolution in an explicit scheme for the time integration.
In some contact scenarios, the constraints undergo large deviations from the initial evaluations at the beginning of time steps, leading to instability with large time steps.
However, in FE simulations, updating the constraints through the constraint resolution is very difficult since it brings about intensive computation costs such as rebuilding the compliance matrix and updating the proximity information.

% contributions
The present work is motivated by providing an implicit scheme for constraint resolution in interactive FE simulations.
Such an implicit scheme can handle the inconsistency of constraint directions at the beginning and the end of time steps.
We propose using a recursive corrective motion scheme to update the constraint directions in Newton iterations.
The constraints could be re-linearized through the iterative method, involving the constraint resolution into an implicit scheme.
To efficiently update the constraint matrices in the recursive scheme, we propose a reformulation of the contact Jacobian matrix, quickly rebuilding the compliance matrix and efficiently computing the boundary state after each correction.
Besides the computational efficiency, our method is also straightforward since it only requires additional matrix multiplication operations, where we could find proven techniques on both the CPU and the GPU.
Moreover, our method remains flexible to be compatible with different collision detection algorithms and resolution methods.

% The rest of this paper is organized as follows. 
% The related works are discussed in Section \ref{sec: related} with respect to the advantages and drawbacks of different methods.
% Section \ref{sec: Background} presents the contact generation, the relevant model of contact and friction, the resolution strategy used in this paper, and the challenges in the current context.
% Section \ref{sec: method recursive corrective motion} and \ref{sec: method updating constraint matrices} are dedicated to the new proposed approach to solve the contact equations, which is finally evaluated Section \ref{sec: results}.

\section{Related works}
\label{sec: related}

% physics-based models: time integration, FEM, PBD, projective dynamics
\subsection{Physics-based models}
Compared to the fast and simple explicit schemes \cite{Comas2008}, implicit time integration scheme \cite{LM_Baraff96} are more suitable in interactive simulations as the external and internal forces are balanced at the end of the time steps.
In multi-object systems, simulating soft solids usually leads to more computational issues than rigid solids due to high degrees of freedom (DOF) and nonlinear mechanics for deformation behavior.
As a gold-standard method, finite element (FE) method is extended in a wide variety of works, such as the linear elastic models \cite{Bro-Nielsen1996}, the co-rotational formulation \cite{Felippa2000}, and the hyperelastic or viscoelastic materials \cite{Marchesseau2010}.
Although providing a good understanding of the deformation mechanisms, FE models remain complex and expensive.

On the other hand, discrete methods such as Position-Based Dynamics (PBD) \cite{Mu2007} provide simple, fast, and robust simulations but suffer from handling the material properties \cite{Bender2014}.
The Projective Dynamics \cite{Bouaziz2014} gives a good trade-off between the performance of PBD and the accuracy of continuum mechanics, addressing the volume conservation problem in PBD.
This work is extended in recent papers such as hyperelastic materials \cite{Liu2017} and frictional contact \cite{ly2020}.
Despite being computationally efficient, the Projective Dynamics remains to validate its capability of simulating realistic materials.
In addition, meshless methods and Neural Networks are other strategies to model soft tissues in real-time \cite{Zhang2018}.

% Time stepping and collision detection
\subsection{Contact generation}
The interactions in contact simulations may lead to discontinuity in the velocities in the mechanical motion.
Generally, two schemes could be used to address this issue:
The \textit{event driven} scheme \cite{Baraff1994} gives a smooth and accurate generation for contacts while the time integration needs to be blocked and the number of instantaneous contacts is restricted.
On the contrary, the \textit{time stepping} scheme \cite{Stewart1996}, \cite{Anitescu2002} involves all potential contact during a fixed time step.
The scheme has no limit on the generated contact and is compatible with both rigid and soft bodies \cite{Saupin08} in implicit time integration.

Searching for the potential contact is commonly performed by the collision detection process.
Fast and straightforward methods may use analytic shapes to approximate the surface of a solid. 
Using the bounding volume hierarchy (BVH), a detailed irregular shape may be represented as a combination of multiple simple shapes \cite{Teschner2005}.
The intersections could be tested with efficient algorithms for common shapes such as sphere and box.
On the other hand, the solid shape could also be represented as polygonal meshes on the surface. 
In a 3D problem, the surface typologies usually consist of triangle elements for rigid and soft solids modeled with FEM.
Testing the intersections could be performed either by the typical way that searches the closest distances between the surface elements or by the advanced methods such as the image-based algorithm using layered depth image (LDI) \cite{Heidelberger2003}.
\cite{Allard2010} extends the work of LDI to GPU implementation and provides a simplification of contact constraints at arbitrary geometry-independent resolution.

The collision detection could be processed either by a discrete or continuous scheme.
Discrete collision detection (DCD) algorithms search the potential contacts in each time step, usually at the beginning.
DCD provides an approximate evaluation of constraint directions and often suffers from the resolution error, which leads to a slight separation or penetration at the end of time steps.
In contrast, continuous collision detection (CCD) algorithms can accurately detect the first time of interaction between solids within a given time step.
A survey of recent works on CCD combined with different detection approaches can be found in \cite{Nie2020}
Nevertheless, with respect to computational cost, CCD is significantly more expensive than DCD, especially in the scenarios with detail shapes for meshes.

% LCP, NCP, Constraint resolution
\subsection{Constraint resolution}

% Lagrange multipliers - constraint-based resolution
The constraint-based methods using Lagrange multipliers (\cite{jean1999235}, \cite{renard2012}) solve the contact problem in a coupled way, providing accurate and robust solutions in contact mechanics for large time steps, where interpenetration is entirely eliminated at the end of time steps.
The Lagrange multipliers methods are commonly formulated as a linear complementarity problem (LCP).
Different numerical methods can be used to address the LCPs in physics-based animations \cite{Erleben2013}.
Direct methods such as pivoting methods give exact resolution, but they are not computationally efficient.
In contrast, iterative methods have been more widely applied in large-scale simulations, especially for those required to perform real-time computations.
Being simple to implement, projected Gauss-Seidel (PGS) (\cite{DDKA06}, \cite{Courtecuisse2009}, \cite{Macklin2016}) can handle the friction response with the Coulomb's friction cone combined in the LCP formulation.
\cite{MACKLIN2019} proposes using a Newton method to solve the non-smooth functions that are reformulated from complementarity problems.

% compliance
To couple contact forces in the resolution and to formulate a system in the \textit{constraint space}, one efficient solution is to formulate the schur-complement of the augmented system, resulting in a compliance matrix (also called \textit{delasus operator}).
Nevertheless, as discussed in \cite{Andrews2021}, the computation of schur-complement tends to be costly when dealing with soft-body since it implies solving a linear system with multiple right-hand sides.
Many methods have been proposed to efficiently process the schur-complement in interactive FE simulations, 
such as the compliance warping technique \cite{Saupin08}, 
the CPU-based parallelization \cite{Schenk2006} and the incomplete factorization \cite{Petra2014} in \textit{Pardiso solver project}, 
the asynchronous precondition-based method \cite{Courtecuisse2010} and the GPU-based parallelization \cite{courtecuisse2014real},
and the updating Cholesky factor in consecutive time steps \cite{Herholz2019} \cite{Herholz2020}.

In different methods for constraint resolution, a common strategy is to linearize the constraints after the collision detection.
The linearization is carried out once in each time step, while the constraint directions may change from the beginning to the end.
The current paper aims to provide a constraint resolution scheme where the constraints could be re-linearized in a recursive method, which is able to handle the change of constraint directions and to bridge the gap between the DCD and CCD.
A fast-updating strategy is proposed to guarantee the efficiency of the recursive scheme, using a straightforward GPU-based implementation.

\section{Background}
\label{sec: Background}

\subsection{Implicit time integration}
For each independent object in a multi-object system, a general description for the physical behavior can be expressed through the Newton's second law:
\begin{equation}
\label{eq:newton's law}
    \mass \acc = \forceextvec - \internalforcefunc (\pos, \vel) + \forceconstraintvec
\end{equation}
where the derivative of the velocity $\acc$ is integrated with the mass matrix $\mass$, the external forces $\forceextvec$, the 
internal forces $\internalforcefunc (\pos, \vel)$, and the constraint forces $\forceconstraintvec$.
To have a balance between different forces at the end of time steps, we choose to integrate the time step $\timestep$ with a backward Euler scheme:
\begin{equation}
\label{eq:implicit integration}
    \vel_{\timestep + \dt} = \vel_{\timestep} + \dt \acc_{\timestep + \dt} \quad
    \pos_{\timestep + \dt} = \pos_{\timestep} + \dt \vel_{\timestep + \dt}
\end{equation}
where $\dt$ is the length of time interval $[\timestep, \timestep + \dt]$.

For rigid solids, as there are no internal forces, the dynamic equation can be simplified as:
\begin{equation}
\label{eq:dynamic equation rigid}
    \mass \Delta \vel_{\timestep + \dt} = \dt \forceextvec + \dt \forceconstraintvec
\end{equation}
with $\Delta \vel=\dt \acc$ is the unknown vector to be solved.

For soft solids, the non-linear function of internal forces is linearized with a first-order Taylor expansion:
\begin{equation}
\label{eq:taylor expansion}
    \internalforcefunc ( \pos_{\timestep + \dt} , \vel_{\timestep + \dt} ) 
    =
    \internalforcefunc (\pos_\timestep, \vel_\timestep)
    + \frac{\partial \internalforcefunc 
    (\pos , \vel)
    }
    {\partial \pos} 
    \dt \vel_{\timestep + \dt}
    + \frac{\partial \internalforcefunc 
    (\pos , \vel)
    }
    {\partial \vel} 
    \dt \acc_{\timestep + \dt}
\end{equation}
In the practice of finite-element simulations, the partial derivative terms are expressed as matrices:
$\frac{\partial \internalforcefunc}{\partial \vel}$ at $(\pos_\timestep, \vel_\timestep)$ as a damping matrix $\damping$,
and $\frac{\partial \internalforcefunc}{\partial \pos}$ at $(\pos_\timestep, \vel_\timestep)$ as a stiffness matrix $\stiffness$. 
The dynamic equation for soft solids finally results in a second order differential equation:
\begin{equation}
\label{eq:dynamic equation soft}
    \underbrace{
    \left[
    \mass + \dt \damping 
    + \dt^2 \stiffness 
    \right]
    }_{ \systemMat}
    % \acc_{\timestep + \dt} 
    \Delta \vel_{\timestep + \dt} 
    = 
    \underbrace{
    \left(
    \dt \forceextvec_{\timestep} 
    -
    \dt \forcevec_{\timestep}
    \right)
    %+ \stiffness \pos_{\timestep} 
    % + \damping \vel_{\timestep} 
    - \dt^2 \stiffness \vel_{\timestep} 
    }_{ \solutionvec}
    + \dt \forceconstraintvec
\end{equation}
with $\forcevec_{\timestep} = \internalforcefunc (\pos_\timestep, \vel_\timestep)$.
For both rigid and soft solids, we have a common formulation of a linear system $\systemMat \Delta \vel = \solutionvec + \dt \forceconstraintvec$ to be solved.

\subsection{Contact generation}

\begin{figure}
     \centering
     \begin{subfigure}[b]{0.4\textwidth}
         \centering
         \includegraphics[width=\textwidth]{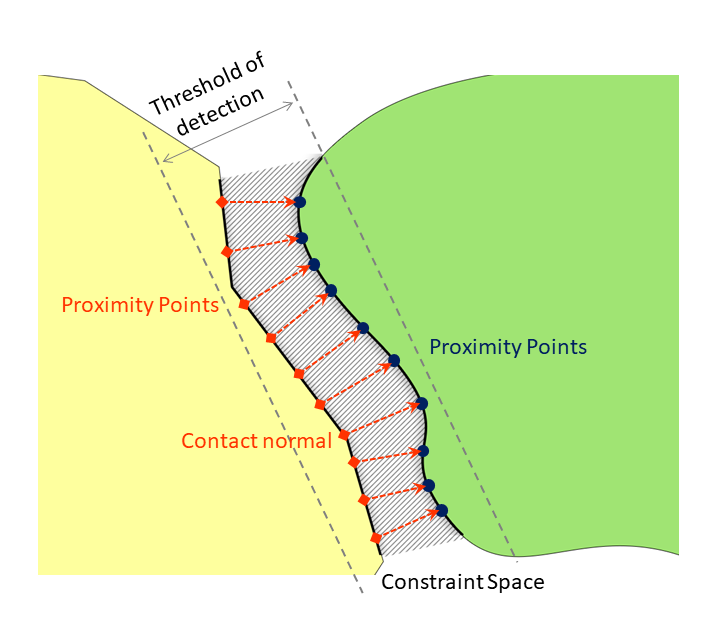}
         \caption{
         For typical method of discrete collision detection, a threshold is usually used to test if the proximity points are close enough.
         The evaluation of potential contact normal depends directly on the surface elements (positions and normals) at the beginning of time steps. 
         }
         \label{fig:initial position based collision detecion}
     \end{subfigure}
     \hspace{2cm}
     \begin{subfigure}[b]{0.4\textwidth}
         \centering
         \includegraphics[width=\textwidth]{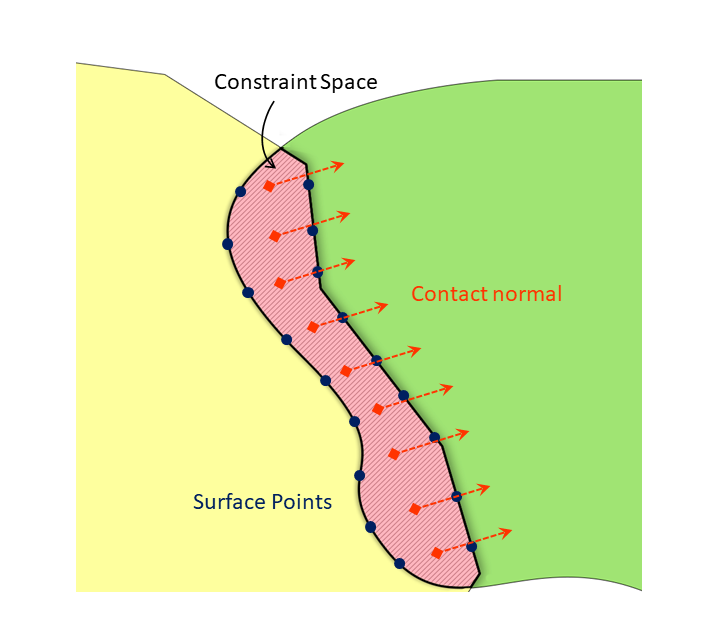}
         \caption{
         For the image-based method of collision detection, the evaluation of contacts is usually based on penetrated volume. 
         The evaluation of constraint normal depends on the boundary of the interpenetration area, which is computed by the positions of surface elements.
         }
         \label{fig:penetration based collision detection}
     \end{subfigure}
        \caption{Constraint linearization with different types of collision detection: 
        To simplify the solving process, collision detection is performed providing a set of discretized constraints between both objects (red arrows).
        Each contact constraint involves the proximity points and a direction of contact normal, which is used to apply the force to separate objects.
        According to different collision detection algorithms, the contact normal is dependent, directly or indirectly, upon the position of proximity points (red and black points on the surface of contacting bodies).}
        \label{fig:collision detection}
\end{figure}

As illustrated in Figure \ref{fig:collision detection}, a contact constraint involves a contact normal and a pair of proximity points that is referred to the objects in touch.
The contact normal defines the direction of a non-interpenetration force to separate the objects.
The pair of proximity points $\poscst$ represent the current state of the surfaces of the objects.
In different methods, $\poscst$ can be directly the points on the object surface, or the representative points using a geometric mapping to transform between $\pos$ and $\poscst$.
Considering the former as an identity mapping, we have a uniformed relationship:
\begin{equation}
\label{eq:geometrcal mapping}
    \poscst = \geomappingfunc(\pos)
\end{equation}
where the function $\geomappingfunc$ is the geometric mapping from the positions of \textit{mechanical DOFs} $\pos$ to the proximity positions $\poscst$.

\subsection{Contact and friction model}
% \subsection{Complementarity}

The contacts between two objects are modeled as constraints, which are discretized and linearized through the collision detection process.
As illustrated in Figure \ref{fig:collision detection}, to prevent interpretations, the distance between two solids $1$ and $2$ can be formulated as a gap function:
\begin{equation}
\label{eq:gap function normal}
    \viol_n(\pos) = \constraintNormal [\poscst_1 - \poscst_2] 
    = \constraintNormal [\geomappingfunc_1(\pos) - \geomappingfunc_2(\pos)] 
    = \mappf_{n 1}(\pos) - \mappf_{n 2}(\pos)
\end{equation}
where the interpretation $\delta_n(\pos)$ is the distance measurement between the proximity positions $\geomappingfunc(\pos)$ projected on the contact normal $\constraintNormal$.
The contact normal is the direction of a force $\fvec_n$ that separates the interpenetrating solids.
The geometrical mapping function $\geomappingfunc(\pos)$ describes the mapping from the \textit{mechanical DOFs} space to the proximity space.
Integrating $\constraintNormal$ into $\geomappingfunc(\pos)$ results in a mapping function $\mappf_n(\pos)$.
Signorini's law presents the complementarity relationship along the constraint direction $\constraintNormal$ for each potential contact:
\begin{equation}
\label{eq:signorini's law}
    0 \leq \viol_n(\pos) \perp \lam_n \geq 0
\end{equation}
where the multiplier $\lam_n$ is the magnitude of the contact force $\fvec_n$ along $\constraintNormal$ as the constraint direction has been normalized.

Equation \eqref{eq:signorini's law} only guarantees to separate the contacting objects.
To model the friction response, the frictional constraints should be added along with the contact normal.
In a 3D problem, a common frictional model complements each contact normal with two tangential directions $\constraintFriction$.
When a contacting is validated ($\lam_n \geq 0$), following Coulomb's friction law, we have:
\begin{equation}
\label{eq:coulomb's law}
% \begin{aligned}
%     \viol_T = 0 \Rightarrow || \forcevec_T || < \lamem || \forcevec_n || \quad (stick)\\
%     \viol_T \neq 0 \Rightarrow  \forcevec_T  = - \lamem || \forcevec_n || \frac{\viol_T}{|| \viol_T ||} \quad (slip)
% \end{aligned}
    0 \leq \viol_f(\pos) \perp \lamem \lam_n - \lam_f \geq 0
\end{equation}
where $\viol_f(\pos)$ is the the interpenetration distance measurement projected on the tangential directions $\constraintFriction$, $\lamem$ is the coefficient of friction, and $\lam_f$ is the value of frictional force along $\constraintFriction$. 
The friction model describes two states for the kinematic behavior: the contacting objects are stuck ($\viol_f = 0$) while $\lam_f \leq  \lamem \lam_n$, and are slipping while $\lam_f$ achieves the maximum value $\lamem \lam_n$.
In addition, $\viol_f(\pos)$ has a similar gap function to Equation \eqref{eq:gap function normal}:
\begin{equation}
\label{eq:gap function friction}
    \viol_f(\pos) = \constraintFriction [\poscst_1 - \poscst_2] 
    = \constraintFriction [\geomappingfunc_1(\pos) - \geomappingfunc_2(\pos)] = \mappf_{f 1}(\pos) - \mappf_{f 2}(\pos_2)
\end{equation}

The governing equations including the complementarity relationships result in the following non-linear system:
\begin{subequations}
\label{eq:KKT}
\begin{empheq}[left=\empheqlbrace]{align}
     &\systemMat_1 \Delta \vel_1 = \solutionvec_1 + \dt \forceconstraintvec_1 \label{eq::KKT1}\\
     &\systemMat_2 \Delta \vel_2 = \solutionvec_2 + \dt \forceconstraintvec_2 \label{eq::KKT2}\\
     &\viol_n(\pos) = \mappf_{n1}(\pos) - \mappf_{n2}(\pos)  \label{eq::KKT3}\\
     &\viol_f(\pos) = \mappf_{f1}(\pos) - \mappf_{f2}(\pos)  \label{eq::KKT4}\\
     &0 \leq \viol_n(\pos) \perp \lam_n \geq 0 \label{eq::KKT5}\\
     &0 \leq \viol_f(\pos) \perp \lamem \lam_n - \lam_f \geq 0 \label{eq::KKT6}
\end{empheq}
\end{subequations}

Since the contact normal constraint along $\constraintNormal$ and the frictional constraint along $\constraintFriction$ have the same mapping function in the gap functions, in practice, the constraints are grouped as constraint sets, where each one of them involves a normal constraint $\constraintNormal$ and two tangential constraints $\constraintFriction$.
Consequently, the functions and vectors can be grouped:
$\mappf_n(\pos)$ and $\mappf_f(\pos)$ are grouped as $\mappf(\pos)$;
$\viol_n(\pos)$ and $\viol_f(\pos)$ are grouped as $\viol(\pos)$;
$\lam_n$ and $\lam_f$ are grouped as $\lam$.

\subsection{Constraint linearization}
% When dealing with the dynamic equation \eqref{eq:dynamic equation soft}, a common strategy is to process a first step called \textit{free motion} that computes the temporary motion $\Delta \vel_{\mathrm{free}}$, which mathematically corresponds to physics dynamics without considering the constraints of contact and friction:
% \begin{equation}
% \label{eq:free motion}
%     \Delta \vel_{\mathrm{free}} = \systemMat^{-1} \solutionvec
% \end{equation}
% A \textit{corrective motion} is later processed to integrate the final motion $\Delta \vel_{\timestep+\dt}$:
% \begin{equation}
% \label{eq:corrective motion}
%     \Delta \vel_{\timestep+\dt} = \Delta \vel_{\mathrm{\mathrm{free}}} + \Delta \vel_{\mathrm{\mathrm{cor}}}
% \end{equation}
% where the corrective term $\Delta \vel^{\mathrm{\mathrm{cor}}}$ is computed according to the unknown constraint forces $\forceconstraintvec$.
When dealing with the dynamic equation \eqref{eq:dynamic equation soft}, a common strategy is to separate the resolution into two motion processes:
\begin{equation}
\label{eq:free motion + corrective motion}
    \Delta \vel_{\timestep+\dt} = 
    \underbrace{
    \systemMat^{-1} \solutionvec 
    }_{\Delta \vel_{\mathrm{free}}}
    +
    \underbrace{
    \dt \systemMat^{-1} \forceconstraintvec
    }_{\Delta \vel_{\mathrm{cor}}}
\end{equation}
where the first integration called \textit{free motion} computes the temporary motion $\Delta \vel_{\mathrm{free}}$, which mathematically corresponds to physics dynamics without considering the constraints of contact and friction.
The second part integrates a \textit{corrective motion} to obtain the motion $\Delta \vel_{\timestep+\dt}$ at the end of time steps.

With the implicit integration (Equation \eqref{eq:implicit integration}), we have the position integration from the intermediate state of \textit{free motion}:
\begin{equation}
\label{eq:position integration}
    \pos_{\timestep + \dt} = \pos_{\mathrm{\mathrm{free}}} + \dt \Delta \vel_{\mathrm{\mathrm{cor}}} 
\end{equation}
where the \textit{corrective motion} could be obtained after the unknown constraint forces $\forceconstraintvec$ are solved: $\Delta \vel_{\mathrm{\mathrm{cor}}} = \systemMat^{-1} \forceconstraintvec$.
The mapping functions $\mappf(\pos)$ are linearized following a first-order Taylor expansion:
\begin{equation}
\label{eq:constraint mapping itegration implicit}
    \mappf (\pos_{\timestep + \dt}) \approx \mappf (\pos_{\mathrm{\mathrm{free}}}) + \dt \frac{\partial \mappf(\pos)}{\partial \pos} \Delta \vel_{\mathrm{\mathrm{cor}}}
\end{equation}
where the Jacobian function $\frac{\partial \mappf(\pos)}{\partial \pos}$ at $\timestep$ the beginning of time steps is usually linearized as the contact Jacobian matrix $\contactJacobian$ that impose the constraints in the mechanical motion of the object.

On the one hand, the contact Jacobian maps \textit{mechanical DOFs} velocities into the velocities for the gap function, allowing to rewrite the gap function in Equation \eqref{eq:gap function normal} and \eqref{eq:gap function friction} as:
\begin{equation}
\label{eq:gap function integrated}
\begin{split}
    \viol^{\timestep + \dt}
    & = \mappf_1(\pos^{\timestep + \dt}_1) - \mappf_2(\pos^{\timestep + \dt}_2) \\
    &\approx 
    \underbrace{
    \mappf_1(\pos^{\mathrm{\mathrm{free}}}_1) - \mappf_2(\pos^{\mathrm{\mathrm{free}}}_2)
    }_{\viol^{\mathrm{\mathrm{free}}} }
    + \dt (\contactJacobian_1 \Delta \vel^{\mathrm{\mathrm{cor}}}_1 - \contactJacobian_2 \Delta \vel^{\mathrm{\mathrm{cor}}}_2)
\end{split}
\end{equation}
where $\viol^{\timestep + \dt}$ and $\viol^{\mathrm{\mathrm{free}}}$ respectively measure the interpenetration at the end of time steps and the state of \textit{free motion}

On the other hand, the contact Jacobian apply the constraint forces into the mechanical motion space:
\begin{equation}
\label{eq:contact force mapping}
    \forceconstraintvec_1 = \contactJacobian_1 \tr \lam \quad \forceconstraintvec_2 = - \contactJacobian_2 \tr \lam
\end{equation}
where the forces are applied for the two objects in opposite directions ($\constraintNormal_1 = - \constraintNormal_2$, $\constraintFriction_1 = - \constraintFriction_2$).
Replace the unknown $\Delta \vel^{\mathrm{\mathrm{cor}}}$ in Equation \eqref{eq:gap function integrated} by Equation \eqref{eq:free motion + corrective motion} and \eqref{eq:contact force mapping}, and we obtain a linear system:
\begin{equation}
\begin{split}
\label{eq:schur complement}
    \viol^{\timestep + \dt}  = 
    \viol^{\mathrm{\mathrm{free}}} +
    \dt^2 
    \underbrace{
    [
    \contactJacobian_1 \systemMat_1^{-1} \contactJacobian_1 \tr + \contactJacobian_2 \systemMat_2^{-1} \contactJacobian_2 \tr 
    ]
    }_{ \delasus} 
    \lam
\end{split}
\end{equation}
where $\delasus \in \realspace^{c \times c}$ (also called \textit{compliance matrix} or \textit{delassus operator} in constrained dynamics) is the schur-complement of the system matrix $\systemMat \in \realspace^{n \times n}$, coupling the constraint forces through the motion space $n$.

% constraint resolution
\subsection{Constraint resolution}
With Equations \eqref{eq:KKT} and \eqref{eq:schur complement}, we have
The governing equations \eqref{eq:KKT} with the schur-complement result in a complementarity system in a constraint space $\realspace^c$:
\begin{subequations}
\label{eq:constraint system}
\begin{empheq}[left=\empheqlbrace]{align}
     &\viol^{\timestep + \dt} = \viol^{\mathrm{\mathrm{free}}} + \dt^2 \delasus \lam \\
     &0 \leq \viol_n^{\timestep + \dt} \perp \lam_n \geq 0 \\
     &0 \leq \viol_f^{\timestep + \dt} \perp \lamem \lam_n - \lam_f \geq 0 
\end{empheq}
\end{subequations}

The non-linear system in Equation \eqref{eq:constraint system} is solved by a projected Gauss-Seidel algorithm \cite{DDKA06} during the successive iterations ($i$):
\begin{equation}
\label{eq:PGS}
\delta_\alpha - \delasus_{\alpha \alpha} \lam_\alpha^{(i)} = \sum_{\beta = 1}^{\alpha-1} \delasus_{\alpha \beta} \lam_\beta^{(i)} + \sum_{\beta = \alpha + 1}^{c} \delasus_{\alpha \beta} \lam_\beta^{(i-1)} + \delta_\alpha^{\mathrm{\mathrm{free}}}
\end{equation}
where $\delasus_{\alpha \beta}$ is a local matrix of $\delasus$ that couples the contact $\alpha$ and $\beta$.
The complementarity problem for each contact group $\alpha$ is solved in the local solution following Signorini's law for the unilateral contact response and Coulomb's law for the frictional response.

Once the $\lam$ is solved, a \textit{corrective motion} is processed to integrate the final motion $\Delta \vel_{\timestep+\dt}$:
\begin{equation}
\label{eq:corrective motion}
\begin{aligned}
    \Delta \vel_1^{\timestep+\dt} = \Delta \vel_1^{\mathrm{\mathrm{free}}} + \dt \systemMat_1^{-1}  \contactJacobian_1 \tr \lam  \\
    \Delta \vel_2^{\timestep+\dt} = \Delta \vel_2^{\mathrm{\mathrm{free}}} - \dt \systemMat_2^{-1}  \contactJacobian_2 \tr \lam
\end{aligned}
\end{equation}

% algorithm of simulation loop
\begin{algorithm}
\SetAlgoLined
 \While{\textit{simulation}}{
 \textit{collision\_detection($\pos^{\timestep}$)} \;
 \textit{constraint\_linearization($\poscst^{\timestep}$)} \;
  \ForEach{$object$}{
  Assemble $\systemMat$, $\solutionvec$, $\contactJacobian$ \;
  $\Delta \vel^{\mathrm{free}} = \systemMat^{-1} \solutionvec$ \;
  $\pos^{\mathrm{free}} = \pos^{\dt} + \dt (\vel^{\dt} + \Delta \vel^{\mathrm{free}})$ \;
 }
 Compute $\viol^{\mathrm{free}}$ according to $\pos^{\mathrm{free}}$\;
 $\delasus = \sum \contactJacobian \systemMat^{-1} \contactJacobian \tr$ \;
 \ForEach{$i \in PGS\_iterations$}{
 \ForEach{$j \in constraint\_groups$}{
 $\viol^c = \viol^{\mathrm{free}} + \delasus \lam^i$ \;
 $\lam^i_j =$ \textit{solve($\lam^i$, $\viol^c$, $\delasus$)} \;
 }
 $\epsilon = \frac{|\lam^i - \lam^{i-1}|}{|\lam^i|}$ \;
 \If{$\epsilon \leq PGS\_error$}{
 \textit{break} \;
 }
 }
 \ForEach{$object$}{
 $\Delta \vel^{\timestep + \dt} = \Delta \vel^{\mathrm{free}} + \dt \systemMat^{-1} \contactJacobian \tr \lam $ \;
 $\vel^{\timestep + \dt} = \vel^{\dt} + \Delta \vel^{\timestep + \dt}$ \;
 $\pos^{\timestep + \dt} = \pos^{\dt} + \dt \vel^{\timestep + \dt}$ \;
 }
 }
   \caption{Standard scheme of simulation loop}
   \label{algo:Standard Simulation Loop}
\end{algorithm}

\section{Recursive corrective motion}
\label{sec: method recursive corrective motion}

\begin{figure}[htb!]
    \centering
\includegraphics[width=\linewidth]{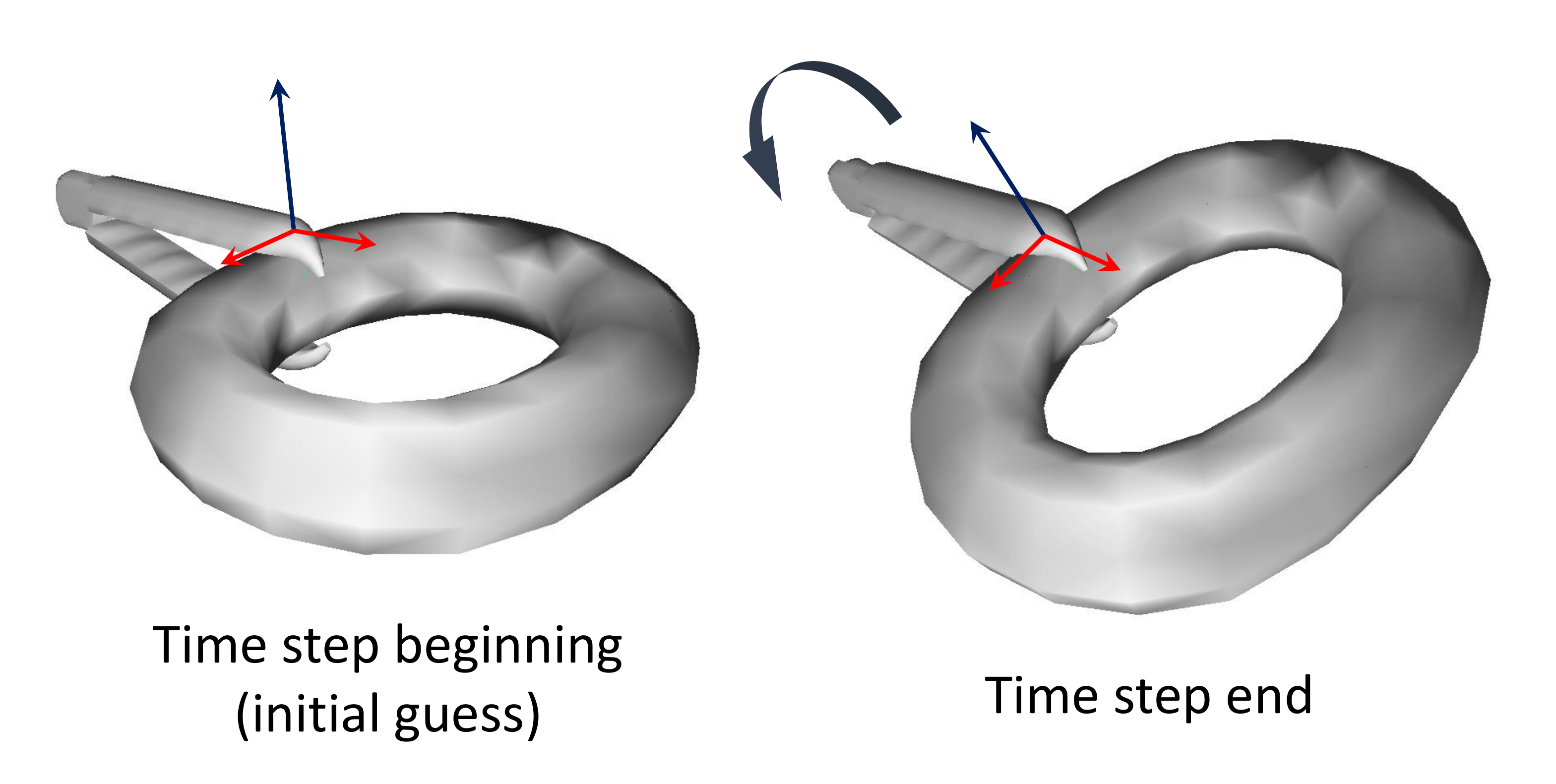}
    \caption{
    Grasping a solid while applying a fast rotation will case a large deviation of constraint directions from the beginning to the end of a time step.
    Such a deviation may cause inaccurate contact responses, instability of contact forces, and eventually failed simulations.
    }
    \label{fig:grasping}
\end{figure}

For both discrete and continuous collision detection algorithms, the constraints will be linearized only once each time step.
Nevertheless, it cannot be guaranteed that the initial evaluation of the constraint directions is always consistent with the state at the end of the time step.
Figure \ref{fig:grasping} gives an example: in a grasping scenario, while acting a moving/rotation, the constraint directions undergo large deviations from the initial guess.

To address this problem, we propose a recursive motion correction scheme:
Instead of using a single corrective motion as in Equation \eqref{eq:free motion + corrective motion}, a iterative correction is applied:
\begin{equation}
\label{eq:recursive motion correction}
    \Delta \vel^{\timestep+\dt} = \Delta \vel^{\mathrm{free}} + (\Delta \vel^{\mathrm{cor}}_1 + \Delta \vel^{\mathrm{cor}}_2 
    + \cdots + \Delta \vel^{\mathrm{cor}}_n)
\end{equation}
where $n$ interactions are performed, and the corrective motion of a given iteration $k$ is computed as following:
\begin{equation}
\label{eq:recursive motion correction 2}
    \Delta \vel^{\mathrm{cor}}_k = \dt \systemMat_k^{-1} \contactJacobian_k \tr \lam_k
\end{equation}
where $\systemMat_k$ and $\contactJacobian_k$ are reevaluated in each iteration according to the current mechanical state ($\pos_k$).
For the integration between two successive iterations $k$ and $k+1$ in Equation \eqref{eq:recursive motion correction}, we have:
\begin{equation}
\label{eq:recursive motion correction 4}
    \pos_{k+1} - \pos_{k} = (\pos^{\mathrm{free}} + \dt \Delta \vel_{k+1}) - (\pos^{\mathrm{free}} + \dt \Delta \vel_k) 
    = \dt \Delta \vel^{\mathrm{cor}}_{k+1}
\end{equation}

Such a recursive scheme is a Newton-Raphson method that provides a more accurate correction.
However, performing such a scheme requires repeating the compliance assembly, the constraint resolution, the corrective motion, and the time integration in recursive iterations.
These additional processes multiply the computational cost, making the system resolution very inefficient.

\begin{algorithm}
\SetAlgoLined
 \ForEach{$k \in Newton\_iterations$}{
    \nl Linearize contact constraints with the proximity positions $\poscst_k$, 
    then update $\contactJacobian_k$ with the constraint directions. \\
    \nl Update compliance matrix: $\delasus_k = \sum \contactJacobian_k \systemMat_k^{-1} \contactJacobian_k \tr$ \\
    \nl Compute constraint forces within local PGS: $\lam = \delasus^{-1} \viol$ \\
    \nl Compute corrective motion: $\Delta \vel^{\mathrm{cor}}_k = \dt \systemMat_k^{-1}  \contactJacobian_k \tr \lam_k $ \\
    \nl Integrate corrective motion: $\vel_k = \vel_{k-1} + \Delta \vel^{\mathrm{cor}}_k$,  $\pos_k = \pos_0 + \dt \vel_k$ \\
    \nl Update the proximity positions: $\poscst_k = \geomappingfunc(\pos_{k})$ \\
    }
   \caption{Recursive correction scheme with standard approach to update the constraint matrices}
   \label{algo: standard recursive correction scheme}
\end{algorithm}

\begin{figure*}[htb!]
    \centering
\includegraphics[width=\linewidth]{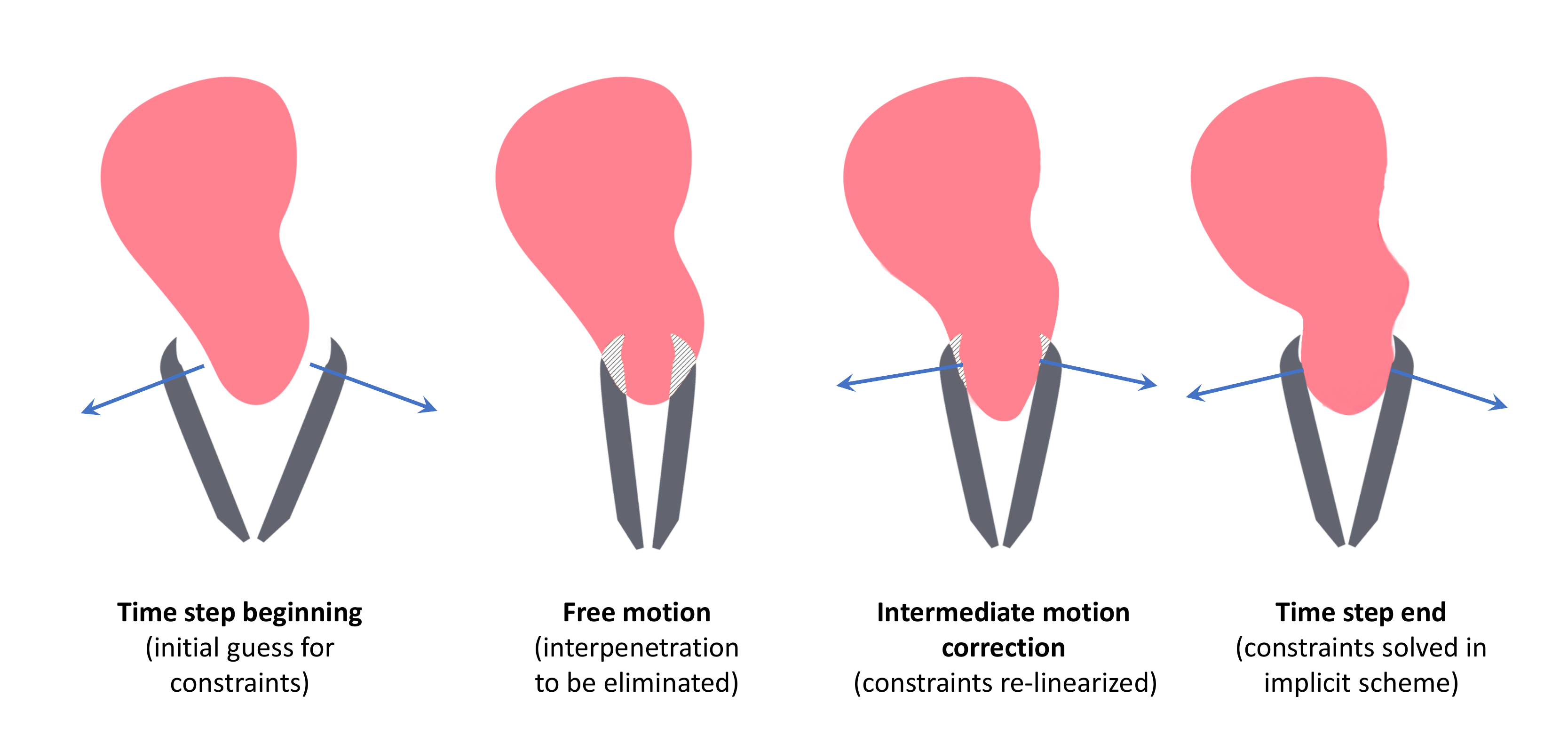}
    \caption{
    Our method solves the constraints in an implicit scheme:
    The constraints are linearized at the beginning of each time step using discrete collision detection.
    Contact and friction responses are computed to eliminate the interpenetration between solids in the free motion.
    The constraint resolution is performed in a recursive corrective motion scheme, where contact forces are computed in each iteration by a local solver.
    Using the contact forces, the boundary conditions of colliding solids and the constraint matrices are updated in the next iteration to compute new motion corrections.
    In this way, the constraint directions are re-linearized recursively until the interpenetration is eliminated.
    }
    \label{fig:recursive scheme}
\end{figure*}

\section{Updating constraint matrices}
\label{sec: method updating constraint matrices}
In this section, we propose a strategy that is able to efficiently process the iterations in Equation \eqref{eq:recursive motion correction}.
In the recursive motion correction scheme (Equation \eqref{eq:recursive motion correction}), 
the system matrix $\systemMat$ is actually the mass matrix $\mass$ for a rigid object. Therefore $\systemMat_k = \systemMat_0 = \mass$, $k \in \realspace^n$.
On the other hand, the corrective motion usually causes small deformation in motion corrections. 
Relying on this hypothesis, we have an approximation for $\systemMat_k$ in the iterations: $\systemMat_k = \systemMat_0$, $k \in \realspace^n$.
Following Equation \eqref{eq:recursive motion correction} and \eqref{eq:recursive motion correction 2}, we have:

\begin{equation}
\label{eq:recursive motion correction 3}
    \Delta \vel^{\timestep+\dt} = \Delta \vel^{\mathrm{free}} + \dt \systemMat^{-1} ( \contactJacobian_0 \tr \lam_0 + \contactJacobian_1 \tr \lam_1 + \cdot \contactJacobian_n \tr \lam_n )
\end{equation}

We recall the definition of the contact Jacobian $\contactJacobian$. 
For a given constraint with a direction $\constraintDirection$, the contribution in $\contactJacobian$ is expressed as:
\begin{equation}
\label{eq:contact jacobian deconstruction}
    \contactJacobian_{(\constraintDirection)} = \frac{\partial \mappf(\pos)}{\partial \pos} 
    = \frac{\partial (\constraintDirection \geomappingfunc(\pos))}{\partial \pos}
     = \constraintDirection \frac{\partial \geomappingfunc(\pos)}{\partial \pos}
\end{equation}
as in practice, the constraint direction $\constraintDirection$ is independent of the mechanical state after a constraint linearization.

With Equation \eqref{eq:contact jacobian deconstruction}, we propose to assembly a Jacobian matrix $\contactJacobianMapping$ for the Jacobian of the geometric mapping $\frac{\partial \geomappingfunc(\pos)}{\partial \pos}$, and a block-diagonal matrix $\contactJacobianDirection$ to store the constraint directions:
\begin{equation}
    \label{eq:jacobian directions}
    \contactJacobianDirection = 
    \begin{bmatrix}
    \constraintDirection_1 & & & \\
     & \constraintDirection_2& & \\
      & &\ddots & \\
       & & & \constraintDirection_c \\
    \end{bmatrix}
\end{equation}
where $\constraintDirection  = \begin{bmatrix} \constraintNormal & \constraintFriction_{(1)} & \constraintFriction_{(2)} \end{bmatrix} $ groups the normal and frictional constraints on a shared proximity point. 
The relation between the two matrices can be actually expressed by a matrix-matrix multiplication:
\begin{equation}
    \contactJacobian = \contactJacobianDirection \contactJacobianMapping 
    \label{eq:isolate directions}
\end{equation}
% illustrate C and H

With Equation \eqref{eq:isolate directions}, a standard compliance assembly (Equation \eqref{eq:schur complement}) can be then reformulated as:
\begin{equation}
    \label{eq:schur complemnt reformulated}
    \delasus = \sum \contactJacobian \systemMat^{-1} \contactJacobian \tr
    =\sum  \contactJacobianDirection 
    \underbrace{\contactJacobianMapping \systemMat^{-1} \contactJacobianMapping \tr
    }_{\delasusMapping}
    \contactJacobianDirection \tr
\end{equation}
where $\delasus$ can be built with $\contactJacobian$ and $\delasusMapping$.
The geometrical mapping $\geomappingfunc(\pos)$ usually undergoes a slight change in the recursive corrective motion scheme.
Based on this hypothesis, $\contactJacobianMapping$ and $\delasusMapping$ are considered invariant during a time step.

Figure \ref{fig:collision detection} illustrates that the linearization of contact constraints will based on the proximity positions $\poscst$. 
With a Taylor expansion, $\poscst$ in the recursive motion correction scheme can be as follows:
\begin{equation}
\label{eq:recursive motion correction 5}
    \poscst_{k+1} = \geomappingfunc(\pos_{k+1})
    \approx \geomappingfunc(\pos_{k}) + \frac{\partial \geomappingfunc(\pos)}{\partial \pos} (\pos_{k+1} - \pos_{k})
\end{equation}
where $k$ and $k+1$ represents two successive iterations in the recursive motion correction scheme.
With Equation \eqref{eq:recursive motion correction 4} and $\frac{\partial \geomappingfunc(\pos)}{\partial \pos}$ linearized as $\contactJacobianMapping$, we continue the development in Equation \eqref{eq:recursive motion correction 5}:
\begin{equation}
\label{eq:recursive motion correction 6}
    \poscst_{k+1}
    \approx \geomappingfunc(\pos_{k}) + \frac{\partial \geomappingfunc(\pos)}{\partial \pos} (\pos_{k+1} - \pos_{k})
    = \poscst_{k} + \dt \contactJacobianMapping \Delta \vel^{\mathrm{cor}}_{k+1}
\end{equation}

Combining Equation \eqref{eq:recursive motion correction 2}, \eqref{eq:recursive motion correction 6}, \eqref{eq:isolate directions} and \eqref{eq:schur complemnt reformulated}, we have:
\begin{equation}
\begin{split}
\label{eq:recursive motion correction 7}
    \poscst_{k+1}
    &\approx \poscst_{k} + \dt \contactJacobianMapping \Delta \vel^{\mathrm{cor}}_{k+1} \\
    &= \poscst_{k} + \dt \contactJacobianMapping (\dt \systemMat^{-1} \contactJacobian_k \tr \lam_k) \\ 
    &= \poscst_{k} + \dt^2 \contactJacobianMapping \systemMat^{-1} \contactJacobianMapping \tr \contactJacobianDirection_k \tr \lam_k \\
    &= \poscst_{k} + \dt^2 \delasusMapping \contactJacobianDirection_k \tr \lam_k
\end{split}
\end{equation}

\pgfplotsset{every axis/.append style={
        scaled y ticks = false, 
        scaled x ticks = false, 
        y tick label style={/pgf/number format/.cd, fixed, fixed zerofill,
                            int detect,1000 sep={\;},precision=3},
        x tick label style={/pgf/number format/.cd, fixed, fixed zerofill,
                            int detect, 1000 sep={},precision=3}
    }
}
 
\begin{table*}[!b]
\setlength{\tabcolsep}{10pt}
\centering
\resizebox{\columnwidth}{!}{
\begin{tabular}{ |c|c||c|c|c|c|c||c|c|c| } 
\hline
\textit{DOFs} & Cst. & Method & Build $\delasusMapping$ & Rebuild $\delasus$ & PGS & Corr. & Newton ite. & Total & Update \\
\hline
\hline
\multirow{2}{*}{18003} & \multirow{2}{*}{308.19} & standard & & 19.89 ms & \multirow{2}{*}{3.76 ms} & 0.82 ms & 24.47 ms & 122.39 ms & 84.60 $\%$  \\
\cline{3-5}\cline{7-10}
& & fast & 19.26 ms & 0.59 ms & & 0.08 ms & 4.43 ms & 41.45 ms & 15.09 $\%$\\
\cline{1-10}
\multirow{2}{*}{24066} & \multirow{2}{*}{365.97} & standard & & 32.21 ms & \multirow{2}{*}{4.84 ms} & 1.19 ms & 38.24 ms & 191.21 ms & 87.34 $\%$  \\
\cline{3-5}\cline{7-10}
& & fast & 31.27 ms & 0.78 ms & & 0.07 ms & 5.69 ms & 59.72 ms & 14.94 $\%$\\
\cline{1-10}
\multirow{2}{*}{30033} & \multirow{2}{*}{410.58} & standard & & 45.47 ms & \multirow{2}{*}{5.87 ms} & 1.47 ms & 52.81 ms & 264.07 ms & 88.88 $\%$ \\
\cline{3-5}\cline{7-10}
& & fast & 45.16 ms & 0.98 ms & & 0.06 ms & 6.91 ms & 79.73 ms & 15.04 $\%$ \\
\cline{1-10}
\multirow{2}{*}{36069} & \multirow{2}{*}{482.25} & standard & & 64.32 ms & \multirow{2}{*}{7.91 ms} & 1.87 ms & 74.10 ms & 370.52 ms & 89.32 $\%$ \\
\cline{3-5}\cline{7-10}
& & fast & 63.73 ms & 1.29 ms & & 0.07 ms & 9.27 ms & 110.10 ms & 14.67 $\%$\\
\hline
\end{tabular}}
\caption{
Comparison of performance between the standard update scheme and the fast update scheme for different numbers of \textit{mechanical DOFs} (\textit{DOFs}) and constraints (Cst.):
For the fast update scheme, building the mapping compliance matrix (Build $\delasusMapping$) is performed once each time step.
The following processes are performed once in each Newton iteration: updating the compliance matrix (Rebuild $\delasus$), processing the local PGS (PGS), and computing the corrective motion (Corr.).
Then the cost of each Newton iteration (Newton ite.) and the whole recursive corrective motion scheme (Total) are compared.
Finally, the proportion of the update constraint matrix process in the Newton iteration (Update) is illustrated.
}
\label{tab:standard vs fast update}
\end{table*}

Now with Equation \eqref{eq:schur complemnt reformulated} and \eqref{eq:recursive motion correction 7}, once $\delasusMapping$ is built, a recursive scheme can be performed in Algorithm \ref{algo: fast recursive correction scheme}.

\begin{algorithm}
\SetAlgoLined
 \ForEach{$k \in Newton\_iterations$}{
    \nl Linearize contact constraints with the proximity positions $\poscst_k$, 
    then update $\contactJacobianDirection_k$ with the constraint directions. \\
    \nl Update compliance matrix: $\delasus_k = \sum \contactJacobianDirection_k \delasusMapping \contactJacobianDirection_k \tr$ \\
    \nl Compute constraint forces within local PGS: $\lam_k = \delasus_k^{-1} \viol$ \\
    \nl Update the proximity positions: $\poscst_k = \poscst_{k-1} + \dt^2 \delasusMapping \contactJacobianDirection_k \tr \lam_k$ \\
    }
   \caption{Recursive correction scheme with fast update of the constraint matrices}
   \label{algo: fast recursive correction scheme}
\end{algorithm}

Compared with the standard update strategy, the computation is greatly simplified in the new scheme.
Instead of inverting the system matrix $\systemMat$, the new scheme only needs matrix-matrix multiplications to update the compliance matrix $\delasus$, and a matrix-vector multiplication to update the boundary state (proximity positions $\poscst$).
As the basic operations in linear algebraic, the matrix multiplications have highly efficient implementations, especially with parallelization on the GPU.

\section{Results and conclusion}
\label{sec: results}

In this section, we evaluate the computation cost of the method presented in Section \ref{sec: method updating constraint matrices} that provides a fast update of constraint matrices in the recursive correction scheme proposed in Section \ref{sec: method recursive corrective motion}.
The simulation tests are conducted in the open-source SOFA framework with a CPU AMD@ Ryzen 9 5950X 16-Core at 3.40GHz with 32GB RAM and a GPU GeForce RTX 3080 10GB.

In order to have an efficient resolution, we use the precondition-based technique in \cite{courtecuisse2014real} to assemble the compliance matrix $\delasus$.
As illustrated in Algorithm \ref{algo: fast recursive correction scheme}, a recursive corrective motion scheme helps to improve the constraint correction in an iterative way, which performs Newton-Rapshon iterations.
Within each Newton iteration, a local \textbf{PGS} is performed to solve the complementarity problem in contact and friction responses, giving temporary contact forces.
The matrix operations in Algorithm \ref{algo: fast recursive correction scheme} are realized with typical GPU-based implementation in \textbf{cuBLAS} and \textbf{cuSPARSE} libraries.

Such contact forces are used to compute the position of the surface elements, which are further used to define the constraint directions in the next iteration.
In Table \ref{tab:standard vs fast update}, we compare the computational cost between the standard scheme and the fast update scheme in 
In a simple contact scenario, a recursive corrective motion scheme with 5 Newton iterations is performed.
Each Newton iteration is accompanied by a local constraint resolution with 30 \textbf{PGS} iterations.
In the standard scheme (see Algorithm \ref{algo: standard recursive correction scheme}), updating constraint matrices takes massive computation cost:
Rebuilding the compliance matrix $\delasus$ and applying the constraint correction with traditional way (see Equation \eqref{eq:corrective motion}) require to invert the system matrix $\systemMat$.
The extra cost by these updating processes is enormous, taking 84-90 $\%$ of each Newton iteration.
On the other hand, in the fast updating scheme (see Algorithm \ref{algo: fast recursive correction scheme}), the computation of the mapping compliance matrix $\delasusMapping$ is outside of the recursive scheme and is performed only once in each time step.
Our method allows fast rebuilding of the compliance matrix (by matrix-matrix multiplications) and efficiently computing the correction on proximity positions (by matrix-vector multiplication).
The matrix multiplication operations are very suitable to be parallelized on GPU.
Moreover, the extra updating cost takes 14-15 $\%$ of each Newton iteration.
Our method is 6.97 $\times$ faster than the standard scheme for each Newton iteration.
For the cost of the whole Newton scheme, including the overhead of building $\delasusMapping$, our method benefits a speedup of 3.20 $\times$ compared to the standard scheme.

Besides the computational performance, our method is straightforward since only matrix operations are required in the recursive scheme.
Moreover, the implementation of the method remains flexible.
In fact, Algorithm \ref{algo: standard recursive correction scheme} could be applied in different ways.
Since the update process is very efficient, it is possible to update the constraint directions in each PGS iteration, or even to carry out the update after the resolution of each constraint, which is similar to the resolution scheme in the Position-Based Dynamics \cite{Mu2007}.

Our future work is to find more applications for the recursive corrective motion scheme with the fast update method.
We hope that our work could inspire readers to help them improve the constraint resolution performance in their works on interactive FE simulations.
% \section{Conclusion}

\footnotesize\smallskip\noindent\textbf{Acknowledgement:} This work was supported by French National Research Agency (ANR) within the project SPERRY ANR-18-CE33-0007.).

\bibliographystyle{ACM-Reference-Format}
\bibliography{content/mimesis}


\begin{thebibliography}{36}

%%% the Web address.

\ifx \showCODEN    \undefined \def \showCODEN     #1{\unskip}     \fi
\ifx \showDOI      \undefined \def \showDOI       #1{#1}\fi
\ifx \showISBNx    \undefined \def \showISBNx     #1{\unskip}     \fi
\ifx \showISBNxiii \undefined \def \showISBNxiii  #1{\unskip}     \fi
\ifx \showISSN     \undefined \def \showISSN      #1{\unskip}     \fi
\ifx \showLCCN     \undefined \def \showLCCN      #1{\unskip}     \fi
\ifx \shownote     \undefined \def \shownote      #1{#1}          \fi
\ifx \showarticletitle \undefined \def \showarticletitle #1{#1}   \fi
\ifx \showURL      \undefined \def \showURL       {\relax}        \fi
% invisible to TeX
\providecommand\bibfield[2]{#2}
\providecommand\bibinfo[2]{#2}
\providecommand\natexlab[1]{#1}
\providecommand\showeprint[2][]{arXiv:#2}

\bibitem[\protect\citeauthoryear{Allard, Faure, Courtecuisse, Falipou, Duriez,
  and Kry}{Allard et~al\mbox{.}}{2010}]%
        {Allard2010}
\bibfield{author}{\bibinfo{person}{J{\'{e}}r{\'{e}}mie Allard},
  \bibinfo{person}{Fran{\c{c}}ois Faure}, \bibinfo{person}{Hadrien
  Courtecuisse}, \bibinfo{person}{Florent Falipou}, \bibinfo{person}{Christian
  Duriez}, {and} \bibinfo{person}{Paul~G. Kry}.}
  \bibinfo{year}{2010}\natexlab{}.
\newblock \showarticletitle{{Volume contact constraints at arbitrary
  resolution}}.
\newblock \bibinfo{journal}{\emph{ACM SIGGRAPH 2010 Papers, SIGGRAPH 2010}}
  \bibinfo{volume}{C}, \bibinfo{number}{3} (\bibinfo{date}{jul}
  \bibinfo{year}{2010}), \bibinfo{pages}{1--10}.
\newblock
\showISBNx{9781450302104}
\showISSN{07300301}
\urldef\tempurl%
\url{https://doi.org/10.1145/1778765.1778819}
\showDOI{\tempurl}


\bibitem[\protect\citeauthoryear{Andrews and Erleben}{Andrews and
  Erleben}{2021}]%
        {Andrews2021}
\bibfield{author}{\bibinfo{person}{Sheldon Andrews} {and}
  \bibinfo{person}{Kenny Erleben}.} \bibinfo{year}{2021}\natexlab{}.
\newblock \showarticletitle{Contact and friction simulation for computer
  graphics}.
\newblock \bibinfo{journal}{\emph{ACM SIGGRAPH 2021 Courses, SIGGRAPH 2021}}
  (\bibinfo{date}{8} \bibinfo{year}{2021}).
\newblock
\showISBNx{9781450383615}
\urldef\tempurl%
\url{https://doi.org/10.1145/3450508.3464571}
\showDOI{\tempurl}


\bibitem[\protect\citeauthoryear{Anitescu and Potra}{Anitescu and
  Potra}{2002}]%
        {Anitescu2002}
\bibfield{author}{\bibinfo{person}{Mihai Anitescu} {and}
  \bibinfo{person}{Florian~A. Potra}.} \bibinfo{year}{2002}\natexlab{}.
\newblock \showarticletitle{A time-stepping method for stiff multibody dynamics
  with contact and friction}.
\newblock \bibinfo{journal}{\emph{Internat. J. Numer. Methods Engrg.}}
  \bibinfo{volume}{55} (\bibinfo{date}{11} \bibinfo{year}{2002}),
  \bibinfo{pages}{753--784}.
\newblock
Issue 7.
\showISSN{1097-0207}
\urldef\tempurl%
\url{https://doi.org/10.1002/NME.512}
\showDOI{\tempurl}


\bibitem[\protect\citeauthoryear{Baksic, Courtecuisse, and Bayle}{Baksic
  et~al\mbox{.}}{2021}]%
        {Baksic2021}
\bibfield{author}{\bibinfo{person}{Paul Baksic}, \bibinfo{person}{Hadrien
  Courtecuisse}, {and} \bibinfo{person}{Bernard Bayle}.}
  \bibinfo{year}{2021}\natexlab{}.
\newblock \showarticletitle{{Shared control strategy for needle insertion into
  deformable tissue using inverse Finite Element simulation}}. In
  \bibinfo{booktitle}{\emph{IEEE International Conference on Robotics and
  Automation}}. \bibinfo{pages}{12442--12448}.
\newblock
\showISBNx{9781728190778}


\bibitem[\protect\citeauthoryear{Baraff}{Baraff}{1994}]%
        {Baraff1994}
\bibfield{author}{\bibinfo{person}{David Baraff}.}
  \bibinfo{year}{1994}\natexlab{}.
\newblock \showarticletitle{Fast contact force computation for nonpenetrating
  rigid bodies}.
\newblock \bibinfo{journal}{\emph{Proceedings of the 21st Annual Conference on
  Computer Graphics and Interactive Techniques, SIGGRAPH 1994}}
  (\bibinfo{date}{7} \bibinfo{year}{1994}), \bibinfo{pages}{23--34}.
\newblock
\showISBNx{0897916670}
\urldef\tempurl%
\url{https://doi.org/10.1145/192161.192168}
\showDOI{\tempurl}


\bibitem[\protect\citeauthoryear{Baraff}{Baraff}{1996}]%
        {LM_Baraff96}
\bibfield{author}{\bibinfo{person}{David Baraff}.}
  \bibinfo{year}{1996}\natexlab{}.
\newblock \showarticletitle{{Linear-time dynamics using Lagrange multipliers}}.
  In \bibinfo{booktitle}{\emph{Proceedings of the 23rd Annual Conference on
  Computer Graphics and Interactive Techniques, SIGGRAPH 1996}}.
  \bibinfo{publisher}{ACM}, \bibinfo{pages}{137--146}.
\newblock
\showISBNx{0897917464}
\urldef\tempurl%
\url{https://doi.org/10.1145/237170.237226}
\showDOI{\tempurl}


\bibitem[\protect\citeauthoryear{Bender, M{\"{u}}ller, Otaduy, Teschner, and
  Macklin}{Bender et~al\mbox{.}}{2014}]%
        {Bender2014}
\bibfield{author}{\bibinfo{person}{Jan Bender}, \bibinfo{person}{Matthias
  M{\"{u}}ller}, \bibinfo{person}{Miguel~A. Otaduy}, \bibinfo{person}{Matthias
  Teschner}, {and} \bibinfo{person}{Miles Macklin}.}
  \bibinfo{year}{2014}\natexlab{}.
\newblock \showarticletitle{{A survey on position-based simulation methods in
  computer graphics}}.
\newblock \bibinfo{journal}{\emph{Computer Graphics Forum}}
  \bibinfo{volume}{33}, \bibinfo{number}{6} (\bibinfo{year}{2014}),
  \bibinfo{pages}{228--251}.
\newblock
\showISSN{14678659}
\urldef\tempurl%
\url{https://doi.org/10.1111/cgf.12346}
\showDOI{\tempurl}


\bibitem[\protect\citeauthoryear{Bouaziz, Martin, Liu, Kavan, and
  Pauly}{Bouaziz et~al\mbox{.}}{2014}]%
        {Bouaziz2014}
\bibfield{author}{\bibinfo{person}{Sofien Bouaziz}, \bibinfo{person}{Sebastian
  Martin}, \bibinfo{person}{Tiantian Liu}, \bibinfo{person}{Ladislav Kavan},
  {and} \bibinfo{person}{Mark Pauly}.} \bibinfo{year}{2014}\natexlab{}.
\newblock \showarticletitle{Projective dynamics}.
\newblock \bibinfo{journal}{\emph{ACM Transactions on Graphics (TOG)}}
  \bibinfo{volume}{33} (\bibinfo{date}{7} \bibinfo{year}{2014}).
\newblock
Issue 4.
\showISSN{15577333}
\urldef\tempurl%
\url{https://doi.org/10.1145/2601097.2601116}
\showDOI{\tempurl}


\bibitem[\protect\citeauthoryear{Bro-Nielsen and Cotin}{Bro-Nielsen and
  Cotin}{1996}]%
        {Bro-Nielsen1996}
\bibfield{author}{\bibinfo{person}{Morten Bro-Nielsen} {and}
  \bibinfo{person}{St{\'{e}}phane Cotin}.} \bibinfo{year}{1996}\natexlab{}.
\newblock \showarticletitle{{Real-time volumetric deformable models for surgery
  simulation using finite elements and condensation}}.
\newblock \bibinfo{journal}{\emph{Computer Graphics Forum}}
  \bibinfo{volume}{15} (\bibinfo{year}{1996}), \bibinfo{pages}{57--66}.
\newblock


\bibitem[\protect\citeauthoryear{Comas, Taylor, Allard, Ourselin, Cotin, and
  Passenger}{Comas et~al\mbox{.}}{2008}]%
        {Comas2008}
\bibfield{author}{\bibinfo{person}{Olivier Comas}, \bibinfo{person}{Zeike~A
  Taylor}, \bibinfo{person}{J{\'{e}}r{\'{e}}mie Allard},
  \bibinfo{person}{S{\'{e}}bastien Ourselin}, \bibinfo{person}{St{\'{e}}phane
  Cotin}, {and} \bibinfo{person}{Josh Passenger}.}
  \bibinfo{year}{2008}\natexlab{}.
\newblock \showarticletitle{{Efficient Nonlinear FEM for Soft Tissue Modelling
  and Its GPU Implementation within the Open Source Framework SOFA}}. In
  \bibinfo{booktitle}{\emph{Biomedical Simulation}}, Vol.~\bibinfo{volume}{5104
  LNCS}. \bibinfo{pages}{28--39}.
\newblock
\showISBNx{3540705201}
\showISSN{03029743}
\urldef\tempurl%
\url{https://doi.org/10.1007/978-3-540-70521-5_4}
\showDOI{\tempurl}


\bibitem[\protect\citeauthoryear{Courtecuisse and Allard}{Courtecuisse and
  Allard}{2009}]%
        {Courtecuisse2009}
\bibfield{author}{\bibinfo{person}{Hadrien Courtecuisse} {and}
  \bibinfo{person}{Jeremie Allard}.} \bibinfo{year}{2009}\natexlab{}.
\newblock \showarticletitle{{Parallel dense gauss-seidel algorithm on many-core
  processors}}. In \bibinfo{booktitle}{\emph{2009 11th IEEE International
  Conference on High Performance Computing and Communications, HPCC 2009}}.
  \bibinfo{publisher}{IEEE}, \bibinfo{pages}{139--147}.
\newblock
\showISBNx{9780769537382}
\urldef\tempurl%
\url{https://doi.org/10.1109/HPCC.2009.51}
\showDOI{\tempurl}


\bibitem[\protect\citeauthoryear{Courtecuisse, Allard, Duriez, and
  Cotin}{Courtecuisse et~al\mbox{.}}{2010}]%
        {Courtecuisse2010}
\bibfield{author}{\bibinfo{person}{Hadrien Courtecuisse},
  \bibinfo{person}{J{\'{e}}r{\'{e}}mie Allard}, \bibinfo{person}{Christian
  Duriez}, {and} \bibinfo{person}{St{\'{e}}phane Cotin}.}
  \bibinfo{year}{2010}\natexlab{}.
\newblock \showarticletitle{{Asynchronous preconditioners for efficient solving
  of non-linear deformations}}. In \bibinfo{booktitle}{\emph{VRIPHYS 2010 - 7th
  Workshop on Virtual Reality Interactions and Physical Simulations}}.
  \bibinfo{pages}{59--68}.
\newblock
\showISBNx{9783905673784}
\urldef\tempurl%
\url{https://doi.org/10.2312/PE/vriphys/vriphys10/059-068}
\showDOI{\tempurl}


\bibitem[\protect\citeauthoryear{Courtecuisse, Allard, Kerfriden, Bordas,
  Cotin, and Duriez}{Courtecuisse et~al\mbox{.}}{2014}]%
        {courtecuisse2014real}
\bibfield{author}{\bibinfo{person}{Hadrien Courtecuisse},
  \bibinfo{person}{J{\'{e}}r{\'{e}}mie Allard}, \bibinfo{person}{Pierre
  Kerfriden}, \bibinfo{person}{St{\'{e}}phane~P.A. Bordas},
  \bibinfo{person}{St{\'{e}}phane Cotin}, {and} \bibinfo{person}{Christian
  Duriez}.} \bibinfo{year}{2014}\natexlab{}.
\newblock \showarticletitle{{Real-time simulation of contact and cutting of
  heterogeneous soft-tissues}}.
\newblock \bibinfo{journal}{\emph{Medical Image Analysis}}
  \bibinfo{volume}{18}, \bibinfo{number}{2} (\bibinfo{year}{2014}),
  \bibinfo{pages}{394--410}.
\newblock
\showISBNx{1361-8415}
\showISSN{13618415}
\urldef\tempurl%
\url{https://doi.org/10.1016/j.media.2013.11.001}
\showDOI{\tempurl}


\bibitem[\protect\citeauthoryear{Duriez}{Duriez}{2013}]%
        {Duriez2013a}
\bibfield{author}{\bibinfo{person}{Christian Duriez}.}
  \bibinfo{year}{2013}\natexlab{}.
\newblock \showarticletitle{{Control of elastic soft robots based on real-time
  finite element method}}.
\newblock \bibinfo{journal}{\emph{Proceedings - IEEE International Conference
  on Robotics and Automation}} (\bibinfo{year}{2013}),
  \bibinfo{pages}{3982--3987}.
\newblock
\showISBNx{9781467356411}
\showISSN{10504729}
\urldef\tempurl%
\url{https://doi.org/10.1109/ICRA.2013.6631138}
\showDOI{\tempurl}


\bibitem[\protect\citeauthoryear{Duriez, Dubois, Kheddar, and Andriot}{Duriez
  et~al\mbox{.}}{2006}]%
        {DDKA06}
\bibfield{author}{\bibinfo{person}{C. Duriez}, \bibinfo{person}{F. Dubois},
  \bibinfo{person}{A. Kheddar}, {and} \bibinfo{person}{C. Andriot}.}
  \bibinfo{year}{2006}\natexlab{}.
\newblock \showarticletitle{{Realistic haptic rendering of interacting
  deformable objects in virtual environments}}.
\newblock \bibinfo{journal}{\emph{IEEE Transactions on Visualization and
  Computer Graphics}} \bibinfo{volume}{12}, \bibinfo{number}{1}
  (\bibinfo{date}{apr} \bibinfo{year}{2006}), \bibinfo{pages}{36--47}.
\newblock
\showISBNx{1077-2626}
\showISSN{10772626}
\urldef\tempurl%
\url{https://doi.org/10.1109/TVCG.2006.13}
\showDOI{\tempurl}
\showeprint[arxiv]{0804.0561}


\bibitem[\protect\citeauthoryear{Erleben}{Erleben}{2013}]%
        {Erleben2013}
\bibfield{author}{\bibinfo{person}{Kenny Erleben}.}
  \bibinfo{year}{2013}\natexlab{}.
\newblock \showarticletitle{{Numerical methods for linear complementarity
  problems in physics-based animation}}.
\newblock \bibinfo{journal}{\emph{ACM SIGGRAPH 2013 Courses, SIGGRAPH 2013}}
  \bibinfo{number}{February} (\bibinfo{year}{2013}).
\newblock
\showISBNx{9781450323390}
\urldef\tempurl%
\url{https://doi.org/10.1145/2504435.2504443}
\showDOI{\tempurl}


\bibitem[\protect\citeauthoryear{Felippa}{Felippa}{2000}]%
        {Felippa2000}
\bibfield{author}{\bibinfo{person}{Ca Felippa}.}
  \bibinfo{year}{2000}\natexlab{}.
\newblock \bibinfo{booktitle}{\emph{{A systematic approach to the
  element-independent corotational dynamics of finite elements}}}.
\newblock \bibinfo{type}{{T}echnical {R}eport} January.
  \bibinfo{institution}{College Of Engineeringuniversity Of Colorado}.
\newblock
\urldef\tempurl%
\url{http://www.colorado.edu/engineering/cas/Felippa.d/FelippaHome.d/Publications.d/Report.CU-CAS-00-03.pdf}
\showURL{%
\tempurl}


\bibitem[\protect\citeauthoryear{Heidelberger, Teschner, and
  Gross}{Heidelberger et~al\mbox{.}}{2003}]%
        {Heidelberger2003}
\bibfield{author}{\bibinfo{person}{Bruno Heidelberger},
  \bibinfo{person}{Matthias Teschner}, {and} \bibinfo{person}{Markus Gross}.}
  \bibinfo{year}{2003}\natexlab{}.
\newblock \showarticletitle{Real-Time Volumetric Intersections of Deforming
  Objects.}
\newblock \bibinfo{journal}{\emph{VMV’03: Proceedings of Vision, Modeling,
  Visualization}}  \bibinfo{volume}{2003}, \bibinfo{pages}{461--468}.
\newblock


\bibitem[\protect\citeauthoryear{Herholz and Alexa}{Herholz and Alexa}{2018}]%
        {Herholz2019}
\bibfield{author}{\bibinfo{person}{Philipp Herholz} {and} \bibinfo{person}{Marc
  Alexa}.} \bibinfo{year}{2018}\natexlab{}.
\newblock \showarticletitle{{Factor once: Reusing Cholesky factorizations on
  sub-meshes}}.
\newblock \bibinfo{journal}{\emph{SIGGRAPH Asia 2018 Technical Papers, SIGGRAPH
  Asia 2018}} \bibinfo{volume}{37}, \bibinfo{number}{6} (\bibinfo{date}{jan}
  \bibinfo{year}{2018}), \bibinfo{pages}{1--9}.
\newblock
\showISBNx{9781450360081}
\showISSN{15577368}
\urldef\tempurl%
\url{https://doi.org/10.1145/3272127.3275107}
\showDOI{\tempurl}


\bibitem[\protect\citeauthoryear{Herholz and Sorkine-Hornung}{Herholz and
  Sorkine-Hornung}{2020}]%
        {Herholz2020}
\bibfield{author}{\bibinfo{person}{Philipp Herholz} {and} \bibinfo{person}{Olga
  Sorkine-Hornung}.} \bibinfo{year}{2020}\natexlab{}.
\newblock \showarticletitle{{Sparse cholesky updates for interactive mesh
  parameterization}}.
\newblock \bibinfo{journal}{\emph{ACM Transactions on Graphics}}
  \bibinfo{volume}{39}, \bibinfo{number}{6} (\bibinfo{year}{2020}).
\newblock
\showISSN{15577368}
\urldef\tempurl%
\url{https://doi.org/10.1145/3414685.3417828}
\showDOI{\tempurl}


\bibitem[\protect\citeauthoryear{Jean}{Jean}{1999}]%
        {jean1999235}
\bibfield{author}{\bibinfo{person}{M. Jean}.} \bibinfo{year}{1999}\natexlab{}.
\newblock \showarticletitle{{The non-smooth contact dynamics method}}.
\newblock \bibinfo{journal}{\emph{Computer Methods in Applied Mechanics and
  Engineering}} \bibinfo{volume}{177}, \bibinfo{number}{3-4}
  (\bibinfo{year}{1999}), \bibinfo{pages}{235--257}.
\newblock
\showISSN{00457825}
\urldef\tempurl%
\url{https://doi.org/10.1016/S0045-7825(98)00383-1}
\showDOI{\tempurl}


\bibitem[\protect\citeauthoryear{Liu, Bouaziz, and Kavan}{Liu
  et~al\mbox{.}}{2017}]%
        {Liu2017}
\bibfield{author}{\bibinfo{person}{Tiantian Liu}, \bibinfo{person}{Sofien
  Bouaziz}, {and} \bibinfo{person}{Ladislav Kavan}.}
  \bibinfo{year}{2017}\natexlab{}.
\newblock \showarticletitle{{Quasi-Newton methods for real-time simulation of
  hyperelastic materials}}.
\newblock \bibinfo{journal}{\emph{ACM Transactions on Graphics}}
  \bibinfo{volume}{36}, \bibinfo{number}{3} (\bibinfo{year}{2017}).
\newblock
\showISSN{15577368}
\urldef\tempurl%
\url{https://doi.org/10.1145/2990496}
\showDOI{\tempurl}


\bibitem[\protect\citeauthoryear{Ly, Jouve, Boissieux, and
  Bertails-Descoubes}{Ly et~al\mbox{.}}{2020}]%
        {ly2020}
\bibfield{author}{\bibinfo{person}{Mickaël Ly}, \bibinfo{person}{Jean Jouve},
  \bibinfo{person}{Laurence Boissieux}, {and} \bibinfo{person}{Florence
  Bertails-Descoubes}.} \bibinfo{year}{2020}\natexlab{}.
\newblock \showarticletitle{Projective dynamics with dry frictional contact}.
\newblock \bibinfo{journal}{\emph{ACM Transactions on Graphics (TOG)}}
  \bibinfo{volume}{39} (\bibinfo{date}{7} \bibinfo{year}{2020}).
\newblock
Issue 4.
\showISSN{15577368}
\urldef\tempurl%
\url{https://doi.org/10.1145/3386569.3392396}
\showDOI{\tempurl}


\bibitem[\protect\citeauthoryear{Macklin, Erleben, M{\"{u}}ller, Chentanez,
  Jeschke, and Makoviychuk}{Macklin et~al\mbox{.}}{2019}]%
        {MACKLIN2019}
\bibfield{author}{\bibinfo{person}{Miles Macklin}, \bibinfo{person}{Kenny
  Erleben}, \bibinfo{person}{Matthias M{\"{u}}ller}, \bibinfo{person}{Nuttapong
  Chentanez}, \bibinfo{person}{Stefan Jeschke}, {and} \bibinfo{person}{Viktor
  Makoviychuk}.} \bibinfo{year}{2019}\natexlab{}.
\newblock \showarticletitle{{Non-smooth Newton methods for deformable
  multi-body dynamics}}.
\newblock \bibinfo{journal}{\emph{ACM Transactions on Graphics}}
  \bibinfo{volume}{38}, \bibinfo{number}{5} (\bibinfo{date}{oct}
  \bibinfo{year}{2019}).
\newblock
\showISSN{15577368}
\urldef\tempurl%
\url{https://doi.org/10.1145/3338695}
\showDOI{\tempurl}
\showeprint[arxiv]{1907.04587}


\bibitem[\protect\citeauthoryear{Macklin, Müller, and Chentanez}{Macklin
  et~al\mbox{.}}{2016}]%
        {Macklin2016}
\bibfield{author}{\bibinfo{person}{Miles Macklin}, \bibinfo{person}{Matthias
  Müller}, {and} \bibinfo{person}{Nuttapong Chentanez}.}
  \bibinfo{year}{2016}\natexlab{}.
\newblock \showarticletitle{XPBD: Position-based simulation of compliant
  constrained dynamics}.
\newblock \bibinfo{journal}{\emph{Proceedings - Motion in Games 2016: 9th
  International Conference on Motion in Games, MIG 2016}} (\bibinfo{date}{10}
  \bibinfo{year}{2016}), \bibinfo{pages}{49--54}.
\newblock
\showISBNx{9781450345927}
\urldef\tempurl%
\url{https://doi.org/10.1145/2994258.2994272}
\showDOI{\tempurl}


\bibitem[\protect\citeauthoryear{Marchesseau, Heimann, Chatelin, Willinger, and
  Delingette}{Marchesseau et~al\mbox{.}}{2010}]%
        {Marchesseau2010}
\bibfield{author}{\bibinfo{person}{St{\'{e}}phanie Marchesseau},
  \bibinfo{person}{Tobias Heimann}, \bibinfo{person}{Simon Chatelin},
  \bibinfo{person}{R{\'{e}}my Willinger}, {and} \bibinfo{person}{Herv{\'{e}}
  Delingette}.} \bibinfo{year}{2010}\natexlab{}.
\newblock \showarticletitle{{Multiplicative Jacobian Energy Decomposition
  Method for Fast Porous Visco-Hyperelastic Soft Tissue Model}}.
\newblock In \bibinfo{booktitle}{\emph{Lecture notes in computer science}}.
  Vol.~\bibinfo{volume}{6361}. \bibinfo{publisher}{Springer},
  \bibinfo{pages}{235--242}.
\newblock
\urldef\tempurl%
\url{https://doi.org/10.1007/978-3-642-15705-9_29}
\showDOI{\tempurl}


\bibitem[\protect\citeauthoryear{Müller, Heidelberger, Hennix, and
  Ratcliff}{Müller et~al\mbox{.}}{2007}]%
        {Mu2007}
\bibfield{author}{\bibinfo{person}{Matthias Müller}, \bibinfo{person}{Bruno
  Heidelberger}, \bibinfo{person}{Marcus Hennix}, {and} \bibinfo{person}{John
  Ratcliff}.} \bibinfo{year}{2007}\natexlab{}.
\newblock \showarticletitle{Position based dynamics}.
\newblock \bibinfo{journal}{\emph{Journal of Visual Communication and Image
  Representation}}  \bibinfo{volume}{18} (\bibinfo{date}{4}
  \bibinfo{year}{2007}), \bibinfo{pages}{109--118}.
\newblock
Issue 2.
\showISSN{10473203}
\urldef\tempurl%
\url{https://doi.org/10.1016/J.JVCIR.2007.01.005}
\showDOI{\tempurl}


\bibitem[\protect\citeauthoryear{Nie, Zhao, Xu, and Li}{Nie
  et~al\mbox{.}}{2020}]%
        {Nie2020}
\bibfield{author}{\bibinfo{person}{Quan Nie}, \bibinfo{person}{Yingfeng Zhao},
  \bibinfo{person}{Li Xu}, {and} \bibinfo{person}{Bin Li}.}
  \bibinfo{year}{2020}\natexlab{}.
\newblock \showarticletitle{A Survey of Continuous Collision Detection}.
\newblock \bibinfo{journal}{\emph{undefined}} (\bibinfo{date}{12}
  \bibinfo{year}{2020}), \bibinfo{pages}{252--257}.
\newblock
\showISBNx{9780738111414}
\urldef\tempurl%
\url{https://doi.org/10.1109/ITCA52113.2020.00061}
\showDOI{\tempurl}


\bibitem[\protect\citeauthoryear{Petra, Schenk, Lubin, and
  G{\"{a}}ertner}{Petra et~al\mbox{.}}{2014}]%
        {Petra2014}
\bibfield{author}{\bibinfo{person}{Cosmin~G. Petra}, \bibinfo{person}{Olaf
  Schenk}, \bibinfo{person}{Miles Lubin}, {and} \bibinfo{person}{Klaus
  G{\"{a}}ertner}.} \bibinfo{year}{2014}\natexlab{}.
\newblock \showarticletitle{{An augmented incomplete factorization approach for
  computing the schur complement in stochastic optimization}}.
\newblock \bibinfo{journal}{\emph{SIAM Journal on Scientific Computing}}
  \bibinfo{volume}{36}, \bibinfo{number}{2} (\bibinfo{year}{2014}),
  \bibinfo{pages}{C139--C162}.
\newblock
\showISSN{10957200}
\urldef\tempurl%
\url{https://doi.org/10.1137/130908737}
\showDOI{\tempurl}


\bibitem[\protect\citeauthoryear{Renard}{Renard}{2013}]%
        {renard2012}
\bibfield{author}{\bibinfo{person}{Yves Renard}.}
  \bibinfo{year}{2013}\natexlab{}.
\newblock \showarticletitle{{Generalized Newton's methods for the approximation
  and resolution of frictional contact problems in elasticity}}.
\newblock \bibinfo{journal}{\emph{Computer Methods in Applied Mechanics and
  Engineering}}  \bibinfo{volume}{256} (\bibinfo{year}{2013}),
  \bibinfo{pages}{38--55}.
\newblock
\showISSN{00457825}
\urldef\tempurl%
\url{https://doi.org/10.1016/j.cma.2012.12.008}
\showDOI{\tempurl}


\bibitem[\protect\citeauthoryear{Saupin, Duriez, Cotin, and Grisoni}{Saupin
  et~al\mbox{.}}{2008}]%
        {Saupin08}
\bibfield{author}{\bibinfo{person}{Guillaume Saupin},
  \bibinfo{person}{Christian Duriez}, \bibinfo{person}{Stephane Cotin}, {and}
  \bibinfo{person}{Laurent Grisoni}.} \bibinfo{year}{2008}\natexlab{}.
\newblock \showarticletitle{Efficient Contact Modeling using Compliance
  Warping}.
\newblock \bibinfo{journal}{\emph{Computer graphics international}}.
\newblock
\urldef\tempurl%
\url{http://hal.inria.fr/hal-00844039}
\showURL{%
\tempurl}


\bibitem[\protect\citeauthoryear{Schenk and G{\"{a}}rtner}{Schenk and
  G{\"{a}}rtner}{2006}]%
        {Schenk2006}
\bibfield{author}{\bibinfo{person}{Olaf Schenk} {and} \bibinfo{person}{Klaus
  G{\"{a}}rtner}.} \bibinfo{year}{2006}\natexlab{}.
\newblock \showarticletitle{{On fast factorization pivoting methods for sparse
  symmetric indefinite systems}}.
\newblock \bibinfo{journal}{\emph{Electronic Transactions on Numerical
  Analysis}}  \bibinfo{volume}{23} (\bibinfo{year}{2006}),
  \bibinfo{pages}{158--179}.
\newblock
\showISBNx{1068-9613}
\showISSN{10689613}


\bibitem[\protect\citeauthoryear{Sifakis and Barbi{\v{c}}}{Sifakis and
  Barbi{\v{c}}}{2012}]%
        {Sifakis2012}
\bibfield{author}{\bibinfo{person}{E Sifakis} {and} \bibinfo{person}{Jernej
  Barbi{\v{c}}}.} \bibinfo{year}{2012}\natexlab{}.
\newblock \showarticletitle{{FEM simulation of 3D deformable solids: a
  practitioner's guide to theory, discretization and model reduction}}.
\newblock \bibinfo{journal}{\emph{ACM SIGGRAPH 2012 Posters}}
  \bibinfo{volume}{85}, \bibinfo{number}{7} (\bibinfo{year}{2012}),
  \bibinfo{pages}{1--35}.
\newblock
\showISBNx{9781450316781}
\showISSN{0385-5414}
\urldef\tempurl%
\url{https://doi.org/10.3987/Contents-12-85-7}
\showDOI{\tempurl}


\bibitem[\protect\citeauthoryear{Stewart and Trinkle}{Stewart and
  Trinkle}{1996}]%
        {Stewart1996}
\bibfield{author}{\bibinfo{person}{D.~E. Stewart} {and} \bibinfo{person}{J.~C.
  Trinkle}.} \bibinfo{year}{1996}\natexlab{}.
\newblock \showarticletitle{{An implicit time-stepping scheme for rigid body
  dynamics with inelastic collisions and coulomb friction}}.
\newblock \bibinfo{journal}{\emph{Internat. J. Numer. Methods Engrg.}}
  \bibinfo{volume}{39}, \bibinfo{number}{15} (\bibinfo{year}{1996}),
  \bibinfo{pages}{2673--2691}.
\newblock
\showISBNx{0-7803-5886-4}
\showISSN{00295981}
\urldef\tempurl%
\url{https://doi.org/10.1002/(SICI)1097-0207(19960815)39:15<2673::AID-NME972>3.0.CO;2-I}
\showDOI{\tempurl}


\bibitem[\protect\citeauthoryear{Teschner, Kimmerle, Heidelberger, Zachmann,
  Raghupathi, Fuhrmann, Cani, Faure, Magnenat-Thalmann, Strasser, and
  Volino}{Teschner et~al\mbox{.}}{2005}]%
        {Teschner2005}
\bibfield{author}{\bibinfo{person}{M. Teschner}, \bibinfo{person}{S. Kimmerle},
  \bibinfo{person}{B. Heidelberger}, \bibinfo{person}{G. Zachmann},
  \bibinfo{person}{L. Raghupathi}, \bibinfo{person}{A. Fuhrmann},
  \bibinfo{person}{M.~P. Cani}, \bibinfo{person}{F. Faure}, \bibinfo{person}{N.
  Magnenat-Thalmann}, \bibinfo{person}{W. Strasser}, {and} \bibinfo{person}{P.
  Volino}.} \bibinfo{year}{2005}\natexlab{}.
\newblock \showarticletitle{Collision Detection for Deformable Objects}.
\newblock \bibinfo{journal}{\emph{Computer Graphics Forum}}
  \bibinfo{volume}{24} (\bibinfo{date}{3} \bibinfo{year}{2005}),
  \bibinfo{pages}{61--81}.
\newblock
Issue 1.
\showISSN{1467-8659}
\urldef\tempurl%
\url{https://doi.org/10.1111/J.1467-8659.2005.00829.X}
\showDOI{\tempurl}


\bibitem[\protect\citeauthoryear{Zhang, Zhong, and Gu}{Zhang
  et~al\mbox{.}}{2018}]%
        {Zhang2018}
\bibfield{author}{\bibinfo{person}{Jinao Zhang}, \bibinfo{person}{Yongmin
  Zhong}, {and} \bibinfo{person}{Chengfan Gu}.}
  \bibinfo{year}{2018}\natexlab{}.
\newblock \showarticletitle{Deformable Models for Surgical Simulation: A
  Survey}.
\newblock \bibinfo{journal}{\emph{IEEE Reviews in Biomedical Engineering}}
  \bibinfo{volume}{11} (\bibinfo{date}{11} \bibinfo{year}{2018}),
  \bibinfo{pages}{143--164}.
\newblock
\showISSN{19411189}
\urldef\tempurl%
\url{https://doi.org/10.1109/RBME.2017.2773521}
\showDOI{\tempurl}


\end{thebibliography}

\end{document}